\documentclass[a4paper]{amsart}
\pdfoutput=1
\usepackage{graphicx}        
\usepackage{multicol}        
\usepackage[bottom]{footmisc}

\usepackage{amsfonts, amsmath, amssymb}
\usepackage{pgf}
\usepackage{subcaption}
\usepackage{multirow}

\usepackage[frozencache=true,cachedir=minted-cache]{minted}
\setminted{breaklines=true, breakanywhere=true}
\definecolor{lbcolor}{rgb}{0.9,0.9,0.9}
\setminted{bgcolor=lbcolor}
\setminted{fontsize=\footnotesize}

\usepackage{mathtools}
\usepackage{booktabs}        
\usepackage{tikz}            
\usetikzlibrary{calc}        
\usetikzlibrary{shapes.geometric}        
\usepackage{csquotes}        
\usepackage[tracking,kerning,spacing]{microtype}

\usepackage[colorinlistoftodos, bordercolor=orange, backgroundcolor=orange!20, linecolor=orange, textsize=scriptsize]{todonotes} 

\usepackage[noend]{algorithmic}

\usepackage{calc,ifthen,alltt,tabularx,capt-of}
\usepackage[linesnumbered,longend,ruled,vlined]{algorithm2e}

\usepackage[
   natbib,
   style=alphabetic,
   sorting=nty,
   maxbibnames=5, 
   maxalphanames=5, 
   isbn=false,url=false]
{biblatex}
\usepackage[bookmarks=false, hypertexnames=false]{hyperref}
\hypersetup{
  colorlinks = true,          
}

\numberwithin{equation}{section}

\usepackage{longtable}
\usepackage{makecell}
\usepackage{wrapfig}

\usepackage[capitalise, noabbrev]{cleveref} 

\newcommand\OSCAR{\texttt{OSCAR}\xspace}
\newcommand\Julia{\texttt{Julia}\xspace}
\newcommand\ANTIC{\texttt{ANTIC}\xspace}
\newcommand\AbstractAlgebra{\texttt{AbstractAlgebra}\xspace}

\newcommand\GAP{\texttt{GAP}\xspace}

\newcommand\Singular{\texttt{Singular}\xspace}

\newcommand\Arb{\texttt{Arb}\xspace}
\newcommand\polymake{\texttt{polymake}\xspace}
\newcommand\polyDB{\texttt{polyDB}\xspace}
\newcommand\mrdi{\texttt{mrdi}\xspace}

\newcommand\cohomcalg{\texttt{cohomCalg}\xspace}
\newcommand\normaliz{\texttt{Normaliz}\xspace}

\newcommand\topcom{\texttt{TOPCOM}\xspace}
\newcommand\mptopcom{\texttt{mptopcom}\xspace}
\newcommand{\cdd}{\texttt{cddlib}\xspace}

\newcommand{\lrs}{\texttt{lrslib}\xspace}

\newcommand\python{\texttt{Python}\xspace}

\newcommand{\SageMath}{\texttt{SageMath}\xspace}

\newcommand\Mathematica{\texttt{Mathematica}\xspace}
\newcommand\Magma{\texttt{Magma}\xspace}
\newcommand\Maple{\texttt{Maple}\xspace}

\newcommand\jupyter{\texttt{Jupyter}\xspace}

\newcommand\Perl{\texttt{Perl}\xspace}

\DeclareUnicodeCharacter{3C0}{$\pi$}
\DeclareUnicodeCharacter{2124}{\ensuremath{\mathbb{Z}}}
\DeclareUnicodeCharacter{2208}{\ensuremath{\in}}
\DeclareUnicodeCharacter{2264}{\ensuremath{\leq}}
\DeclareUnicodeCharacter{1D456}{\ensuremath{i}}

\definecolor{promptColor}{rgb}{0.0,0.0,0.589}
\definecolor{brkpromptColor}{rgb}{0.589,0.0,0.0}
\definecolor{gapinputColor}{rgb}{0.589,0.0,0.0}
\definecolor{gapoutputColor}{rgb}{0.0,0.0,0.0}
\definecolor{darkgreen}{rgb}{0.05,0.6,0.1}
\definecolor{colrem}{rgb}{0,0.7,0}

{\bfseries}{\itshape}

\newcommand{\tmfloatcontents}{}
\newlength{\tmfloatwidth}
\newcommand{\tmfloat}[5]{
	\renewcommand{\tmfloatcontents}{#4}
	\setlength{\tmfloatwidth}{\widthof{\tmfloatcontents}+1in}
	\ifthenelse{\equal{#2}{small}}
	{\setlength{\tmfloatwidth}{0.45\linewidth}}
	{\setlength{\tmfloatwidth}{\linewidth}}
	\begin{minipage}[#1]{\tmfloatwidth}
		\begin{center}
			\tmfloatcontents
			\captionof{#3}{#5}
		\end{center}
\end{minipage}}

\newcounter{truefigure}
\makeatletter
\renewcommand{\p@subfigure}{}
\makeatother

\newcommand{\PreserveBackslash}[1]{\let\temp=\\#1\let\\=\temp}
\newcolumntype{C}[1]{>{\PreserveBackslash\centering$}p{#1}<{$}}
\newcolumntype{R}[1]{>{\PreserveBackslash\raggedleft$}p{#1}<{$}}
\newcolumntype{L}[1]{>{\PreserveBackslash\raggedright$}p{#1}<{$}}

{\bfseries}{\rmfamily}

\ifx\tocless\undefined
  \newcommand{\tocless}[1]{#1}
\fi

\title{Confirmable Workflows in \OSCAR}
\author{Michael Joswig}
\author{Lars Kastner}
\author{Benjamin Lorenz}
\address[Michael Joswig]{TU Berlin, Chair of Discrete Mathematics/Geometry, Berlin, Germany; Max-Planck-Institut für Mathematik in den Naturwissenschaften, Leipzig, Germany, \url{joswig@math.tu-berlin.de}}
\address[Lars Kastner]{TU Berlin, Chair of Discrete Mathematics/Geometry, Berlin, Germany, \url{kastner@math.tu-berlin.de}}
\address[Benjamin Lorenz]{TU Berlin, Chair of Discrete Mathematics/Geometry, Berlin, Germany, \url{lorenz@math.tu-berlin.de}}

\begin{document}
\begin{abstract}
  We discuss what is special about the reproducibility of workflows in computer
   algebra. It is emphasized how the programming language \Julia and the new
   computer algebra system \OSCAR support such a reproducibility, and how users
   can benefit for their own work.
\end{abstract}
\maketitle


\section{Introduction}
\label{ch:sp-joswig-kastner-lorenz-confirmable-workflows}
\label{confirmable:intro}

There is a current trend to rethink the role of and the use of data in all areas of scientific research.
Some aspects are common to all fields of science, such as questions concerning, e.g., where to store the data to make it long-term accessible to the scientific community, or the role of metadata, which should help finding specific data sets.
But some other aspects also vary drastically from field to field.
Most importantly, the fundamental question about which kind of data is worth storing and how.
The primary goals are described in the FAIR-principles, which mean that data should be Findable, Accessible, Interoperable and Reproducible.
Here we look at research data in mathematics, and computer algebra in particular, from the angle of reproducibility.
On the way we will also talk about data interoperability, but only to a lesser degree.
Throughout we will give examples based on computations with the computer algebra system \OSCAR, which is written in \Julia.
These examples are chosen to highlight the specific features which the pair \Julia/\OSCAR has to offer toward the reproducibility of computer algebra experiments.

Computer experiments have formed an invaluable resource in mathematical research since the advent of computers.
Large calculations that are infeasible for humans with pen and paper can be completed in seconds.
A hypothesis can be tested on a million examples.
However these advantages can only unfold their true potential if the calculation in the computer is not regarded as a black box.
Instead a computer experiment needs to be accompanied by documentation and supplementary material such that everyone reading these details is convinced of the correctness of the calculation.
Here we are restricting our attention to computer algebra rather than, say, numerical analysis.
This distinction means that, in our scenario, computations are always exact, and questions concerning the precision or the quality of an approximation, do not arise.
Furthermore, data sets in computer algebra often tend to be smaller than their counter-parts in numerical analysis.
In contrast, computer algebra data tends to be more varied, and equipped with more intricate semantics.

It is a main goal of this chapter to sketch guidelines for the proper documentation of a computer experiment with \OSCAR, such that the results can easily be confirmed by other mathematicians, with \OSCAR or without.
This confirmation has two steps:
First there is the confirmation that re-running the experiment leads to an equivalent result.
Second there is the mathematical soundness that the experiment provides the mathematical insight claimed.
Along the way we discuss design decisions and other choices.

\tocless{\subsection{Related Work}}
\noindent
There is an ongoing effort to discuss the reliability and the reproducibility of research results in all areas of science.
In Germany this led to setting up a large scientific network under the name \emph{Nationale Forschungsdateninitiative (NFDI)} (National Research Data Initiative).
The \emph{Mathematical Research Data Initiative (MaRDI)} is the single consortium within the NFDI which addresses research data in the mathematical sciences.
Recently, MaRDI published a white paper \cite{mardi:whitepaper} on research
data management in mathematics.

The MaRDI recommendations are derived from a substantial body of previous work.
For example, Stodden et al. \cite{stodden_306_study} published a survey on 306 articles from the Journal of Computational Physics.
They found that more than half of the results were impossible to reproduce.
The authors emphasized that none of the results could be reproduced with minimal effort.
One of their main conclusions was that many problems could have been avoided if the principles of \cite{bbs:facitlitating_reproducibility} had been followed.

A more recent example is a study of Samuel and Mietchen \cite{mietchen:jupyter} of almost $10\,000$ \jupyter notebooks from biomedical publications.
That study focuses on \python notebooks.
In the end, less than 10\% of the notebooks could be rerun successfully.
Even fewer produced the same results as originally.

An article by Riedel et al.\ \cite{Riedel_study}
\enquote{analyzes the reproducibility of 108 publications from an interdisciplinary Collaborative Research Center on applied mathematics in various scientific fields.}
To summarize the conclusion: Only about 40\% of the publications considered gave enough data for attempting replication at all.
Out of these, only 4 were considered \enquote{fully reproducible}.

A comprehensive study of reproducibility in computer algebra seems to be lacking so far.
However, we invite the reader to try to pick a random computer algebra publication from 10 years ago and to reproduce these results.
Our own experiments suggest that the reader will most likely experience some problems.
The project \texttt{MathRepo} \cite{MathRepo} seeks to improve this situation.

\section{The FAIR Principles}\label{section:fair}
The Paris Call on Research Assessment\footnote{\url{https://osec2022.eu/paris-call/}} establishes the following principles for main conditions concerning research data in general:
\begin{quote}
  Ensure independence and transparency of the data, infrastructure and criteria
  necessary for research assessment and for determining research impacts; in particular
  by clear and transparent data collection, algorithms and indicators, by ensuring control
  and ownership by the research community over critical infrastructures and tools, and
  by allowing those assessed to have access to the data, analyses and criteria used.
\end{quote}
These principles are also reflected in the San Francisco Declaration on Research Assessment\footnote{\url{https://sfdora.org/}}.
Requirements for data to be useful to other scientists is often translated into the \emph{FAIR principles} \cite{fair}.
Here F stands for \enquote{findable}, A for \enquote{accessible}, I for \enquote{interoperable}, and R for \enquote{reusable}.
This means that a researcher who is working in the field that the data belongs to will find it without prior knowledge, she can actually get at the data, and use it interoperable in other contexts.
Reusability entails that the data can be reproduced.
Only the latter property turns a data set a scientifically meaningful contribution.

That being said, it is not entirely obvious what it requires to call a data set in computer algebra FAIR.
Here we focus on the reusability/reproducibility and briefly mention interoperability in computer algebra.
That is to say, we largely ignore the important aspects of findability and accessibility.
As we see it, reproducibility is the outcome of an entire workflow, which is confirmable, i.e., adhering to the FAIR principles, at any point.
This has a wide range of implications.
For instance, this requires that the software code producing the data needs to be FAIR itself.
This clearly invites the use of open source software.
Conversely, closed software is maybe not ruled out entirely, but the reproducibility of computations with current commercial software systems would require a level of service by the manufacturers of these systems, which seems to be hard to find.

It is worth noting that the reproducibility of proofs in mathematics may be a challenge, independent from computers. 
A prominent example is the classification of the finite simple groups, which was accepted as accomplished by many around 1980, but some questions remained.
Consequently, Gorenstein, Lyons and Solomon initiated a revised classification in a sequence of books \cite{Finite+Simple+Groups}, which is still growing.
A major outcome of this classification and, at the same time, also an indispensable tool along the way, is the \emph{Atlas Of Finite Groups} \cite{CCN85}.
The question of its reproducibility was raised by Jean-Pierre Serre, when he asked for proofs of the Atlas tables during a talk in 2015\footnote{Serre's presentation can be viewed on YouTube at \url{https://youtu.be/MZ6_JKYdKog}; the Atlas is mentioned after 35min, approximately.}.
In \cite{BMO17} Breuer, Malle and O'Brien made an effort towards the verification of the Atlas, with \GAP and \Magma.
So this is an example where mathematical software has been used for checking proofs which had been obtained by more traditional ways of doing mathematics.

Ultimately, any issues concerning the correctness of mathematical proofs, computational or not, should be resolved by theorem provers like \texttt{Coq}\footnote{\url{https://coq.inria.fr/}} or \texttt{LEAN}\footnote{\url{https://leanprover-community.github.io/}}.
Despite tremendous progress, these provers are still far from capable of certifying large computer algebra computations.
Some algorithms can be certified this way.
However, typically neither an actual implementation nor its results (considered coming from a black box) can be verified with \texttt{Coq}, \texttt{LEAN} or their friends.
This discrepancy leaves a gap, which needs to be filled with a computational workflow which can be confirmed.


\section{Computer Algebra Experiments}

A confirmable workflow essentially entails two things: First, the entire workflow must be reproducible.
Second, its theoretical claims need to be verifiable by other researchers.
A typical scenario is the review process of an article which contains claims which are proved computationally.
In the discussion of related work, we gave references to existing principles and guidelines.
The purpose of this section is to put these guidelines in the context of computer algebra.

One particularity of computer algebra is its fragmentation.
There are many niche software packages solving only very special problems in a highly optimized fashion.
Examples include \cohomcalg\footnote{\url{http://wwwth.mppmu.mpg.de/members/blumenha/cohomcalg/}} \cite{Blumenhagen:2010pv} for computing the cohomology of toric varieties, \topcom \cite{topcom} and \mptopcom \cite{mptopcom} for enumerating triangulations of finite point sets, or \Arb \cite{Johansson2017arb} for arbitrary precision interval arithmetic.
This list could be continued nearly indefinitely.
If one needs many of these, the resulting workflow may become complicated.
Consequently, it is impossible to provide a scheme for a confirmable workflow in computer algebra which is too rigid.
The discussion of general purpose computer algebra systems will be deferred to Section~\ref{section:oscar}.
Since the developers of these systems provide a coherent view on the components used, the situation becomes somewhat easier for users of general purpose computer algebra systems.
However, even in this case, it is necessary to take into account that some libraries may be linked dynamically, which thus depend on the local installation.

In the rest of this section we give explicit recommendations toward confirmability.

\subsection{Connect Your Experiment to the Mathematical Theory}\label{section:tasks:mathematical_theory}
Especially in mathematics and in computer algebra, objects and the theory behind them may be quite complicated.
At the same time the computation may seem easy, like simply enumerating lattice points in a polytope or computing the kernel of a matrix.
It is essential to explain how the computation fits into the theory.
The detailed steps are the items 1, 2 and 3 in \cite[Section 9.2.1]{bbs:facitlitating_reproducibility}.

Linking computations with the underlying theory can be achieved in several ways:
A very simple approach suggests to add code snippets to a written article or, conversely, references to that article as comments in the code.
It is a drawback that this method tends to get messy fast, and it is hard to maintain in large projects.
A second approach suggests using a \jupyter notebook or similar as an intermediate layer between the code and the paper.
Another key part is the proper documentation of the code, especially if it more than just a script, e.g., a \Julia package.

The same remarks hold for any data, too.
Often data looks deceptively simple, like matrices with integer or rational entries or a mere collection of finite sets.
For this data to be usable in the future it needs to be thoroughly connected to the mathematical context.
For instance, if a matrix is supposed to describe some polytope, then it needs to be specified explicitly if the rows or the columns are supposed to form the generators of an interior description, or if the rows or the columns should be interpreted as linear or affine inequalities.

\subsection{Publish Your Data}\label{section:tasks:data}
To be able to replicate your experiment any reader needs to know the input and output of your program or code, see item 10 in \cite[Section 9.2.1]{bbs:facitlitating_reproducibility}.
In particular, it is not enough to publish the code only and then let the reader figure out the input for your examples.
If your experiment produces intermediate results, it makes sense to store these intermediate results too.
This is especially important when the intermediate result is the output of one software which is passed on as the input to another software component.
Intermediate results provide a level of redundancy which is desirable.
They help to understand the full workflow, and they also help to detect the precise point of failure, when things should go wrong.

It is a subtle point that most current computer algebra systems do not provide a high-level view on the data produced with them.
For instance, \SageMath makes use of \python's \emph{pickle} system, which is fragile because it relies on the memory layout of the data during runtime.
\SageMath sometimes compensates such pickles by also storing code snippets which can ideally be used for the reproduction.
This setup poses a challenge for the long-term maintenance of data sets.
As detailed below, \OSCAR follows a different strategy, by employing the new \mrdi file format \cite{mrdi-file-format}.

Data formats in computer algebra are still being established and probably will
not be unified in the near future. Thus documentation will need to be done very
carefully and diligently. As mentioned in the previous section, the data stored
may actually be simple in structure, e.g. consisting of matrices or vectors.
However the mathematical object encoded can be much more complex. In order for
the data to be FAIR, see Section~\ref{section:fair}, every dataset needs to be published
with backlinks to the mathematical theory.

An important side aspect is item 13 of \cite[Section 9.2.1]{bbs:facitlitating_reproducibility}, choosing an appropriate license.
There are tools and questionnaires for finding a suitable license\footnote{\url{https://opendefinition.org/licenses/}}.
So let us just highlight the following: In most areas of the world, data without a license cannot be used for further research, legally.

\subsection{Publish Your Code}\label{section:tasks:code}
Your code is the key ingredient to rerunning your experiment. There are
different ways to publish it, for example on the arXiv you can upload your code
as an auxiliary file. This approach is only suitable in case your code is very
small, say a \jupyter notebook or a simple script file. In most cases though
hosting your code in a public git repository is the gold standard. This can be
on GitHub or a similar platform. Many universities offer GitLab instances for
publishing code. To guarantee long term availability of the code, sign it up
for mirroring by software heritage\footnote{\url{https://www.softwareheritage.org}}.

\subsection{Write Down All Software Versions}\label{section:tasks:software_versions}
Usually one will not write the entire algorithm from scratch, but instead rely
on algorithms implemented by other mathematicians, experts in their respective
fields. 

Not every setup will produce the same results. Here by \enquote{the same} we mean
really the exact same digital data, bit by bit the same. Depending on your
setup, data might be represented differently internally. Then algorithms that
run in parallel might not produce output in the same order. Finally the
mathematical definition of things being equal often differs from the much more
strict notion of \enquote{digital equality}. Besides recording these mathematical
details carefully, one should also document details about the setup used. Not
all aspects apply to every experiment, so apply common sense, and when in
doubt, record whichever thing was causing you doubts.

Changes to the underlying software may change the behavior of your code. The
best case is of course that your code still does the same. The second best case
is that you get an error message, because some function has been renamed or
some other part of your code now fails. The worst case is a silent change of
behavior, undetected by your tests, but nevertheless eliminating mathematical
soundness.

Here is one small example.  The following is valid code in all versions of \Julia:
\begin{minted}{jl}
  x = [1,2]
  x[2], x[1] = x
\end{minted}
Yet, the behavior in \Julia version 1.6 is very different from the behavior in version 1.7 and onward.
To analyze and understand such a small example is not difficult, but issues like the above can make it difficult to analyze larger chunks of code.
Such challenges, which arise from semantic changes during the evolution of a programming language, are quite common.

We conclude this section with another pitfall which is sometimes overlooked.
Most executables (on a Linux system anyway) are dynamically linked to system libraries.
They may be moving parts, too.
The following command in a Linux shell shows the dependencies of the \Julia interpreter itself.
\begin{minted}{bash}
[lars@nomurai ~]$ ldd `which julia-1.9.3`
	linux-vdso.so.1 (0x00007ffd14f34000)
	libdl.so.2 => /usr/lib/libdl.so.2 (0x00007f12c07f0000)
	libpthread.so.0 => /usr/lib/libpthread.so.0 (0x00007f12c07eb000)
	libc.so.6 => /usr/lib/libc.so.6 (0x00007f12c0609000)
	/lib64/ld-linux-x86-64.so.2 => /usr/lib64/ld-linux-x86-64.so.2 (0x00007f12c0836000)
	libjulia.so.1 => not found
\end{minted}

\subsection{Document Your Hardware and Operating System}\label{section:tasks:hardware}

This section is especially important when measuring the performance of algorithms, but it can also help explain to some reader, why your experiment was running out of memory on their machine, but not on yours.
Additionally different architectures may influence the results in subtle ways, meaning that you may get different results on a Mac vs a Windows machine, e.g., if the relevant algorithms employ randomization.
In that case you should of course make sure that these results remain the same mathematically speaking.

On Linux, a tool like \mintinline{bash}{inxi} can help collect information about CPU,
memory and storage.
\begin{minted}{bash}
[lars@nomurai ~]$ inxi -a
CPU: quad core 11th Gen Intel Core i7-1165G7 (-MT MCP-)
speed/min/max: 745/400/4700 MHz Kernel: 5.16.20-200.fc35.x86_64 x86_64
Up: 17d 16h 41m Mem: 11804.1/31819.6 MiB (37.1%)
Storage: 1.86 TiB (29.2% used) Procs: 373 Shell: Bash 5.1.8 inxi: 3.3.14
\end{minted}
You can get basic information about your operating system using \mintinline{bash}{uname}:
\begin{minted}{bash}
[lars@nomurai ~]$ uname -a
Linux nomurai 5.16.20-200.fc35.x86_64 #1 SMP PREEMPT Wed Apr 13 22:09:20 UTC 2022 x86_64 x86_64 x86_64 GNU/Linux
\end{minted}
For Linux systems it is useful to also add information about the Linux distribution.

\section{\Julia and Her Friends}
\OSCAR relies on Julia as a programming language to combine the cornerstones
into a unified computer algebra system.  In addition to that, the Julia
environment provides several key features that support confirmable workflows.

\subsection{The Package Manager}\label{section:julia:package_manager}
The package manager \texttt{Pkg.jl} is integrated as a core package into the \Julia
distribution and can be accessed directly from the \Julia prompt. This allows
for quick installations and does not require any extra permissions.

All installed packages are stored in a content-addressed manner in a
user-specific folder. This way multiple versions of the same package can be
installed at the same time even though only one may be loaded in a single
session. But this allows users to work on different projects which might require
different dependencies.
Content-addressed refers to the fact that each version of a package is stored
using an identifier that is based on a \emph{tree hash}, a combined hash
of all files and folders in that package directory.

From a developer perspective, the packaging system is also very easy to use and
well integrated into GitHub to allow easy registration of new packages and
version updates into the \emph{general registry}, the main package
database that \texttt{Pkg.jl} uses. It is open to anyone, provided the package is covered
by an open license, follows reasonable naming guidelines and semantic
versioning.
The tree hash is also used as an identifier to mirror all registered package
versions from their repository to the \Julia package servers. This makes sure
that the packages stay available independent of their original repositories.

\subsection{Project Environments}
\label{sec:TOML}
A key feature for confirmable workflows is that the package manager is built around project environments, these consist of two TOML files to describe all packages that are used for a given project.
TOML is short for \enquote{Tom's Obvious Minimal Language}\footnote{\url{https://toml.io/en/}}.

\begin{figure}
\begin{minted}{js}
name = "Oscar"
uuid = "f1435218-dba5-11e9-1e4d-f1a5fab5fc13"
authors = ["The OSCAR Team <oscar@mathematik.uni-kl.de>"]
version = "1.0.0"

[deps]
AbstractAlgebra = "c3fe647b-3220-5bb0-a1ea-a7954cac585d"
AlgebraicSolving = "66b61cbe-0446-4d5d-9090-1ff510639f9d"
Distributed = "8ba89e20-285c-5b6f-9357-94700520ee1b"
DocStringExtensions = "ffbed154-4ef7-542d-bbb7-c09d3a79fcae"
GAP = "c863536a-3901-11e9-33e7-d5cd0df7b904"
Hecke = "3e1990a7-5d81-5526-99ce-9ba3ff248f21"
JSON = "682c06a0-de6a-54ab-a142-c8b1cf79cde6"
JSON3 = "0f8b85d8-7281-11e9-16c2-39a750bddbf1"
LazyArtifacts = "4af54fe1-eca0-43a8-85a7-787d91b784e3"
Nemo = "2edaba10-b0f1-5616-af89-8c11ac63239a"
Pkg = "44cfe95a-1eb2-52ea-b672-e2afdf69b78f"
Polymake = "d720cf60-89b5-51f5-aff5-213f193123e7"
Preferences = "21216c6a-2e73-6563-6e65-726566657250"
Random = "9a3f8284-a2c9-5f02-9a11-845980a1fd5c"
RandomExtensions = "fb686558-2515-59ef-acaa-46db3789a887"
Serialization = "9e88b42a-f829-5b0c-bbe9-9e923198166b"
Singular = "bcd08a7b-43d2-5ff7-b6d4-c458787f915c"
TOPCOM_jll = "36f60fef-b880-50dc-9289-4aaecee93cc3"
UUIDs = "cf7118a7-6976-5b1a-9a39-7adc72f591a4"
cohomCalg_jll = "5558cf25-a90e-53b0-b813-cadaa3ae7ade"

[compat]
AbstractAlgebra = "0.40.1"
AlgebraicSolving = "0.4.11"
Distributed = "1.6"
DocStringExtensions = "0.8, 0.9"
GAP = "0.10.2"
Hecke = "0.30.0"
JSON = "^0.20, ^0.21"
JSON3 = "1.13.2"
LazyArtifacts = "1.6"
Nemo = "0.43.0"
Pkg = "1.6"
Polymake = "0.11.14"
Preferences = "1"
Random = "1.6"
RandomExtensions = "0.4.3"
Serialization = "1.6"
Singular = "0.22.4"
TOPCOM_jll = "0.17.8"
UUIDs = "1.6"
cohomCalg_jll = "0.32.0"
julia = "1.6"
\end{minted}
  \caption{Project.toml file of \texttt{Oscar.jl}}
  \label{fig:Project.toml}          
\end{figure}

\begin{description}
   \item[\texttt{Project.toml}] The first one is the project file which lists all direct
      dependencies for the project, together with optional compatibility bounds.
      Any \Julia package contains such a project file, in this case it will also
      contain the name, version, and a unique identifier for that package.  But
      it may also be used for an unnamed project which only lists other packages
      that are in use by this project.
      See Figure~\ref{fig:Project.toml} for \OSCAR's own \texttt{Project.toml}.
   \item[\texttt{Manifest.toml}] The second file describes a concrete instantiation of
      such a project file and contains the full list of all direct and indirect
      dependencies together with their exact version numbers and tree hashes.
\end{description}

Given a project file, the package manager can be used to create or update the
manifest to allow installation of all packages for a given project.
But from a reproducibility standpoint the key feature is that given a project file
together with a manifest, the package manager can also be used to reinstall
the precise versions that were used when this manifest was created. This
can be extremely useful to recreate software experiments.

The currently active manifest can be inspected in the \texttt{Pkg.jl} REPL.
Even for a \Julia core package like \texttt{Downloads.jl} the manifest already
contains 12 entries.
\begin{minted}{julia}
(project) pkg> status --manifest
Status `~/project/Manifest.toml`
  [0dad84c5] ArgTools v1.1.1
  [56f22d72] Artifacts
  [f43a241f] Downloads v1.6.0
  [7b1f6079] FileWatching
  [b27032c2] LibCURL v0.6.4
  [8f399da3] Libdl
  [ca575930] NetworkOptions v1.2.0
  [deac9b47] LibCURL_jll v8.4.0+0
  [29816b5a] LibSSH2_jll v1.11.0+1
  [c8ffd9c3] MbedTLS_jll v2.28.2+0
  [14a3606d] MozillaCACerts_jll v2022.10.11
  [83775a58] Zlib_jll v1.2.13+0
  [8e850ede] nghttp2_jll v1.52.0+1
\end{minted}

An established approach to document and store computer experiments in computer
algebra are notebooks, and here \jupyter is the open source standard.
While a \jupyter notebook in general does not suffice to fully document the
setup \cite{mietchen:jupyter}, it can be augmented with \Julia's project
environments to make sure the correct dependencies are installed automatically.

The combination of the two TOML files can be seen as \Julia's
contribution to a predefined software environment, meaning a software
environment that can reproduce the exact same setting a computer experiment was
previously run in. 
A computer experiment with a more intricate setup can be made reproducible with a suitable
\emph{MaPS}\footnote{\url{https://github.com/aaruni96/maps}} runtime environment
or other
containerization solution, like Docker\footnote{\url{https://www.docker.com/}}
or even a VM. \emph{MaPS} provides a dedicated runtime
\begin{center}
   \texttt{org.oscar\_system.oscar/x86\_64/1.0.0}
\end{center}
for using \OSCAR version 1.0, in particular, for running the examples from the
upcoming \OSCAR book \cite{OSCAR-book}.

\subsection{Binary Packages}
In addition to \Julia packages \texttt{Pkg.jl} also manages binary packages
for non-julia software that may be used by \Julia packages. This works by having
a small \Julia package in the general registry that refers to a set of
pre-compiled binaries for all supported host platforms.

In particular, this means that the project files, including the manifest, not
only fix specific versions for \Julia packages, but these will also be used to
install all required binary packages in the correct versions.

All binary packages for \Julia are built using
\texttt{BinaryBuilder.jl}\footnote{\url{https://binarybuilder.org/}} and the
build recipes are managed in the Yggdrasil
repository\footnote{\url{https://github.com/JuliaPackaging/Yggdrasil}}.  The
build system also includes several mechanisms to allow the binaries to be
reproducible, i.e. rebuilding a package from the same recipe will in many cases
produce bit-identical output.

\subsection{GitHub Integration}
The package managing process for \Julia is well integrated into GitHub, we want
to highlight a few important features provided by various \Julia packages:

\begin{itemize}
   \item Automatic registration of new packages and new package versions with a
      simple comment. This allows for a quick turnaround for bug fixes and new
      features. And it also ensures that all versions are properly registered,
      together with a tree hash, and stay available for download.
   \item Each package version can automatically generate a new GitHub release
      which can then be imported to Zenodo to create a citable DOI for the
      package.
   \item There are several helpers for the free GitHub Actions continuous
      integration system which makes it very easy to automatically run tests
      during the development, even on different platforms that might not be
      available to the developers.
\end{itemize}

\section{\OSCAR}\label{section:oscar}

In this section we discuss features of \OSCAR which support a confirmable workflow.
First, we give hints to how users can organize their own \OSCAR experiments such that they become reproducible by other researchers.
Second, we address code documentation and maintenance of \OSCAR itself as well as the serialization of data.
Knowing about these more technical aspects will help to further improve the documentation of user experiments.

\subsection{Documenting Computational Experiments}
\Julia and \OSCAR offer a variety of tools to aid in documentation of computer experiments of varying size.
Here we will discuss how these tools support the user in the tasks of sections \ref{section:tasks:mathematical_theory} to \ref{section:tasks:software_versions}.

As described in section~\ref{section:tasks:mathematical_theory}, the code should be connected to the surrounding mathematical theory.
This can be done via a \jupyter notebook for small projects, or via some custom documentation using \texttt{Documenter.jl} as described in section \ref{section:oscar:user_manual}.

Data can be published according to section~\ref{section:tasks:data} using \OSCAR's serializer \cite{mrdi-file-format}, see section~\ref{section:oscar:serialization} for more details.
One vision of the MaRDI project is to establish this file format in the community such that other computer algebra systems can read data from \OSCAR and vice versa, thereby increasing interoperability.

Publishing the code as in section~\ref{section:tasks:code} can be done using any suitable repository, see for example \cite{mardi:whitepaper} for guidance.
Again, for smaller projects a notebook may be enough, for a larger project with intended reuse it makes sense to think about creating a separate \Julia module and adding it to the \Julia package manager, see section~\ref{section:julia:package_manager}.

If the project results in code that are interesting to a part of the \OSCAR community there is a dedicated process for adding new functionality to \OSCAR\footnote{\url{https://docs.oscar-system.org/dev/Experimental/intro/}}.

Lastly, both \OSCAR and \Julia provide convenient \mintinline{julia}{versioninfo} commands for collecting the relevant software versions of section~\ref{section:tasks:software_versions}.
The Oscar variant also takes several additional parameters to give an even fuller view of the versions installed:\label{confirmable:versioninfo}
\begin{minted}{jlcon}
julia> Oscar.versioninfo(full=true)
OSCAR version 1.0.0
  combining:
    AbstractAlgebra.jl   v0.40.1
    GAP.jl               v0.10.3
    Hecke.jl             v0.30.2
    Nemo.jl              v0.43.1
    Polymake.jl          v0.11.14
    Singular.jl          v0.22.4
  building on:
    Antic_jll               v0.201.500+0
    Arb_jll                 v200.2300.0+0
    Calcium_jll             v0.401.100+0
    FLINT_jll               v200.900.9+0
    GAP_jll                 v400.1200.200+9
    Singular_jll            v403.214.1400+0
    libpolymake_julia_jll   v0.11.4+0
    libsingular_julia_jll   v0.43.0+0
    polymake_jll            v400.1100.1+0
See `]st -m` for a full list of dependencies.

Julia Version 1.10.2
Commit bd47eca2c8a (2024-03-01 10:14 UTC)
Build Info:
  Official https://julialang.org/ release
Platform Info:
  OS: Linux (x86_64-linux-gnu)
  CPU: 8 × 11th Gen Intel(R) Core(TM) i7-1165G7 @ 2.80GHz
  WORD_SIZE: 64
  LIBM: libopenlibm
  LLVM: libLLVM-15.0.7 (ORCJIT, tigerlake)
Threads: 1 default, 0 interactive, 1 GC (on 8 virtual cores)
Official https://julialang.org/ release
\end{minted}
You can see that this also entails information on the OS and the CPU.
Furthermore it has information on the dependencies and even their version numbers.
This version data is also stored in the TOML files mentioned in Section~\ref{sec:TOML}.

\subsection{User Manual}\label{section:oscar:user_manual}
The entire \OSCAR user manual resides next to the actual code, in the \OSCAR source tree.
In fact, it comprises separate sections dedicated to specific areas of mathematics.
Each such area starts with an introductory page which explains the basic concepts and definitions.
In each section we list standard mathematical textbooks which are used as the basis for all functions in that section.

For instance, toric varieties are algebraic varieties which are defined via a polyhedral fan associated with a lattice polytope.
In the literature, two competing conventions exist, which are dual to one another.
Either the fan is the normal fan of the relevant polytope, or the face fan.
The textbook \cite{CLS:2011} by Cox, Little and Schenck is chosen as \OSCAR's standard reference, and so toric varieties in \OSCAR are defined via normal fans.

\OSCAR's functions are documented in the \Julia code, and we rely on \texttt{Documenter.jl} to compile the various pieces of the documentation into HTML.
Those HTML pages are linked on \OSCAR's GitHub page.
They exist in two flavors: one for the latest stable release\footnote{\url{https://docs.oscar-system.org/stable/}} and one for the current developer's version\footnote{\url{https://docs.oscar-system.org/dev/}}.
Further versions, e.g., for git branches, can be built locally by calling \mintinline{jl}{Oscar.build_doc()}.
The code examples that accompany the documentation are verified via \mintinline{jl}{Documenter.jldoctest()} as part of the continuous integration process; see Section~\ref{sec:confirmable:CI}.

\subsection{Data Serialization}\label{section:oscar:serialization}
Serialization in computer science is the process of taking a chunk of data and putting it into a form such that it can be written to a file and read back identically.
Sometimes writing to a file is replaced by network or interprocess communication.
Between these tasks there are differences that matter on a technical level, but there is no difference between the communication over a network and writing to and reading from a file on a logical level.
Principally, \texttt{Serialization.jl} takes care of the serialization of objects in \Julia, and here is the beginning of the documentation of its main function \mintinline{jl}{serialize}:
\begin{quote}
  Write an arbitrary value to a stream in an opaque format, such that it can be read back by deserialize.
  The read-back value will be as identical as possible to the original, but note that Ptr values are serialized as all-zero bit patterns (NULL).
\end{quote}
The little caveat at the end of that text fragment gives a first hint that serialization is sometimes not as straightforward as it might seem.
This is particularly true in semantically rich environments like in computer algebra.

The traditional way of storing data resulting from computer algebra computations is via notebooks, which combine code, computational results and text. 
This idea was pioneered by \Mathematica and later picked up by \Maple and others.
\Julia and thus \OSCAR is supported by \jupyter, and so such notebooks exist for \OSCAR, too.
The notebook concept comes with serious limitations, however.
Namely, it is difficult or impossible to store intermediate results without the need to recompute them when they are needed for a subsequent new computation.
In this way \Mathematica, \Maple and \jupyter behave almost the same.

\OSCAR employs the new \mrdi file format \cite{mrdi-file-format} for the serialization of its data.
The design goals are different from \texttt{Serialization.jl} or from notebooks.
In \OSCAR, the user should be able to start a computation on one machine, store the resulting output in a file, read that file on a different machine and continue that computation, without the need of recomputation.
Of course, it is trivial to print output and to store it in a text file.
For computer algebra this typically does not suffice, for several reasons.
First, parsing that text before continuing the computation may be tedious.
Second, the same may be even impossible since the usual text output sometimes suppresses crucial information.
Third, what if between the two steps of the computation several years have passed, and the specific way of printing a certain output has changed?
Ideally, such a serialization could even pave the road to a certain degree of interoperability among several computer algebra systems.
However, this is a challenge of its own.
We refer to the article \cite{mrdi-file-format} for answers to these questions and more technical details.

To conclude this section we give one example, to show what \OSCAR can do, and which would be impossible with \texttt{Serialization.jl}.
We create a multivariate polynomial ring with coefficients in the finite field of order 49.
\begin{minted}{jlcon}
julia> F, o = finite_field(7,2)
(Finite field of degree 2 and characteristic 7, o)

julia> R, (y,z) = F["y","z"]
(Multivariate polynomial ring in 2 variables over GF(7, 2), FqMPolyRingElem[y, z])
\end{minted}
\noindent Then we save two polynomials that are loaded back and added.
\begin{minted}{jlcon}
julia> save("p.mrdi", 2*y^3*z^4 + (o + 3)*z^2 + 5*o*y + 1)

julia> save("q.mrdi", y^17*z^3 + (o + 4)*z + 2*o*y)

julia> print( load("p.mrdi") + load("q.mrdi") )
y^17*z^3 + 2*y^3*z^4 + (o + 3)*z^2 + (o + 4)*z + 1
\end{minted}

The key point here is that the semantics of \OSCAR allows for adding two polynomials only if they live in the same ring; it is not enough that the two polynomials are both bivariate over the same domain of coefficients.
Those coefficients pose a further problem; since 49 is not a prime number there is no canonical encoding for the coefficients.
Again \OSCAR semantics requires the coefficients to live in the same field.
To sketch the range of applications that are possible with the \mrdi file format, let us imagine the following scenario.
After storing the polynomials to the files \mintinline{jl}{p.mrdi} and \mintinline{jl}{q.mrdi}, the user, Alice, could send those files via email to her friends Bob and Charlie.
Then Bob computes the square of \mintinline{jl}{p.mrdi} on his computer, while Charlie computes the third power \mintinline{jl}{q.mrdi}.
Both store their results in new files.
Afterwards they send their files back to Alice, who is able to compute the sum of that square and that third power.
Bob and Charlie do not even need to know about each other.

\subsection{Confirmable \OSCAR Book}
As a demonstration we have prepared a workflow that runs all the examples from
the different chapters of this book. This is done via a specialized package
called \texttt{OscarBookExamples.jl}\footnote{\url{https://github.com/oscar-system/OscarBookExamples.jl}}.
Let us summarize the steps to do so:
\begin{enumerate}
   \item For every chapter, we read the \LaTeX\xspace code and extract the lines
      including code files. Then we read those code files in order and compose
      them into markdown files.
   \item These markdown files are then added to the documentation part of the module
      \texttt{OscarBookExamples.jl} and the \Julia package
      \texttt{Documenter.jl} is run on this documentation. We set it to
      automatically correct the output in the examples which are comprised of
      the examples from the \OSCAR book.
   \item The corrected examples are extracted from the markdown files and
      compared to the original examples. In case there are differences, these
      are reported and can then be fixed by the authors.
\end{enumerate}
We use several mechanisms from \texttt{Documenter.jl}.
All examples are wrapped
in \texttt{jldoctest} environments. For every chapter a custom label is
introduced and all \texttt{jldoctest}s are equipped with this label. This
ensures that these examples are run in the same environment and that variables
may be reused.

Some examples use code that is not in \OSCAR, this code has been placed
separate directory \texttt{auxiliary\_code}.

Already during the writing phase, the continuous testing of the code included
exhibited many shortcomings of the examples:
\begin{enumerate}
   \item Typos in the code.
   \item Examples use code that has not been merged to master yet.
   \item Examples use external code that is not up to \OSCAR standards.
\end{enumerate}

We collected the examples from this book in the \texttt{OscarBookExamples.jl} repository.

\subsection{Continuous Integration}
\label{sec:confirmable:CI}
\OSCAR utilizes a multitude of software components written in several languages, including \texttt{C}, \texttt{C++}, \Perl, \python and, of course, \Julia.
The largest contributions come from the four \emph{cornerstone systems}: \ANTIC, \GAP, \polymake and \Singular.
These in turn interface other software; e.g., \polymake will often dispatch convex hull computations to \cdd, \lrs, or \normaliz.
What makes \OSCAR special is that the development team of \OSCAR comprises the developers of the four cornerstones.
This results in an unusually high degree of integration of each cornerstone into \OSCAR, leading to a coherent semantics and fast computations.

This peculiar setup requires a comprehensive approach to testing and maintaining the code, to avoid any regression, with respect to correctness and performance, during the development process.
To this end \OSCAR employs the continuous integration (CI) workflow of GitHub, including code coverage analysis. This aids the developer in properly checking all new contributions.
Additionally, the cornerstones continue to exist as separate software systems, which thus come with their own CI features and setups.
It is useful that GitHub's CI workflow directly allows for continuous delivery (CD).
Such a setup for CI/CD is is considered quite standard by now and similarly
implemented for many other computer algebra systems.
These tools are crucial for guaranteeing robustness in a complex system like \OSCAR with many moving parts, while keeping development speed high.

For instance, duplicate implementations of standard algorithms can be used for cross-certification. Here is a Smith normal form computation:
\begin{minted}{jlcon}
julia> m = matrix(ZZ, [1 2; 3 4])
[1   2]
[3   4]

julia> snf(m)
[1   0]
[0   2]
\end{minted}
\noindent \OSCAR attaches multiple implementations of Smith normal forms to the same function, for instance an optimized version for integer matrices, which is the one used above. A second one from \AbstractAlgebra works for arbitrary Euclidean rings.
\begin{minted}{jlcon}
julia> methods(snf, Tuple{MatrixElem{ZZRingElem}})
# 2 methods for generic function "snf" from AbstractAlgebra:
 [1] snf(x::ZZMatrix)
     @ Nemo ~/.julia/packages/Nemo/xwSoX/src/flint/fmpz_mat.jl:1096
 [2] snf(a::MatrixElem{T}) where T<:RingElement
     @ ~/.julia/packages/AbstractAlgebra/dsta0/src/Matrix.jl:5431

julia> invoke(snf, Tuple{MatrixElem{ZZRingElem}}, m) == snf(m)
true
\end{minted}
\OSCAR uses sensible defaults for the user through \Julia's multiple dispatch system which is why invoking the more general \AbstractAlgebra implementation requires a syntactic effort.

\section{Conclusion and Outlook}
\OSCAR was created with the FAIR principles in mind. It
allows various users to document their computer experiments,
and in particular, saving intermediate results. 
Already now \OSCAR is able to interact with the database \polyDB\cite{polyDB}, which contains objects from polyhedral and tropical geometry. We plan to setup further databases of a similar kind. Modern document based databases support JSON, which underlies the \mrdi file format. In this way our approach to serialization naturally works with existing database solutions.

Ideally, computer algebra data would last forever, much like theorems in mathematics. This is in contrast with existing software that needs to evolve and adapt to new standards and hardware.
Similarly, data files need to evolve, too. 
In \OSCAR we rely on versioning and automatic upgrade scripts, based on two decades of experience with the \polymake file format~\cite{polymake_XML:ICMS_2016}, see also~\cite{MMTCM:eg-models}.

We envision interacting with computer algebra systems other than \OSCAR through the \mrdi file format and namespaces.

We integrated the collected examples of the upcoming \OSCAR book
\cite{OSCAR-book} into the \OSCAR CI to continuously monitor whether the
examples still work with the most recent \OSCAR version. This will help us to
migrate the existing example code to future versions of \OSCAR.

\subsection*{Acknowledgements}
Michael Joswig and Lars Kastner were supported by MaRDI (Mathematical Research
Data Initiative), funded by the Deutsche Forschungsgemeinschaft (DFG), project
number 460135501, NFDI 29/1 ``MaRDI -- Mathematische
Forschungsdateninitiative''\footnote{\url{https://www.mardi4nfdi.de/}}.
The authors wish to thank Thomas Breuer for pointing out and detailing the
pitfalls and struggles while verifying the Atlas tables.
Michael Joswig is supported by the SFB-TRR 195--286237555 ``Symbolic Tools in
Mathematics and their
Application''\footnote{\url{https://www.computeralgebra.de/sfb/}}.

\printbibliography
\end{document}